\documentclass[useAMS,usenatbib]{mn2e}

\usepackage{enumerate}
\usepackage{graphicx}
\usepackage{natbib}
\usepackage{longtable}
\usepackage{rotating}
\usepackage[usenames,dvipsnames]{color}
\usepackage{hyperref}
\usepackage{lineno}

\newcommand{\Rmnum}[1]{\uppercase\expandafter{\romannumeral#1}}

\title[COMs formation on Stochastically Heated Grains]
{Complex Organic Molecules Formation in Cold Cores on Stochastically Heated Grains}
\author[Chen et al.]{
Long-Fei Chen$^{1,2}$\thanks{chenlongf@nao.cas.cn}, 
Qiang Chang$^{3}$, 
Yao Wang$^{4}$, 
Di Li$^{2,5,6}$\thanks{dili@nao.cas.cn}\\
$^{1}$Research Center for Intelligent Computing Platforms, Zhejiang Laboratory, Hangzhou 311100, China\\
$^{2}$National Astronomical Observatories, Chinese Academy of Sciences, 20A Datun Road, Chaoyang District, Beijing 100101, China\\
$^{3}$School of Physics and Optoelectronic Engineering, Shandong University of Technology, Zibo 255000, China\\
$^{4}$Purple Mountain Observatory and Key Laboratory of Radio Astronomy, Chinese Academy of Sciences, 10 Yuanhua Road, Nanjing 210023, China\\
$^{5}$University of Chinese Academy of Sciences, Beijing 100049, China\\
$^{6}$NAOC-UKZN Computational Astrophysics Centre, University of KwaZulu-Natal, Durban 4000, South Africa\\
}

\date{Accepted 2022 September 6. Received 2022 August 31; in original form 2022 June 26}

\begin{document}

\pagerange{\pageref{firstpage}--\pageref{lastpage}} \pubyear{2022}

\maketitle

\label{firstpage}

\begin{abstract}
We investigate the roles of stochastic grain heating in the formation of complex organic molecules (COMs) in cold cores, where COMs have been detected.
Two different types of grain-size distributions are used in the chemical models. 
The first one is the MRN distribution, 
and the second one considers grain coagulation to study its effects on the chemical evolution in these environments.
The macroscopic Monte Carlo method is used to perform the two-phase chemical model simulations.
We find that (1) grain coagulation can affect certain gas-phase species, such as CO$_2$ and N$_2$H$^+$, in the cold core environments, which can be attributed to the volatile precursors originating from the small grains with temperature fluctuations;
(2) grains with radii around 4.6 $\times$ 10$^{-3}$ $\mu$m contribute most to the production of COMs on dust grains
under cold core conditions, while few species can be formed on even smaller grains with radii less than 2 $\times$ 10$^{-3}$ $\mu$m;
(3) COMs formed on stochastically heated grains could help explain the observed abundances of gas-phase COMs in cold cores.
\end{abstract}

\begin{keywords}
astrochemistry -- ISM: abundances -- ISM: molecules
\end{keywords}

%correcting citations, verbs........

\section{Introduction}
Interstellar complex organic molecules (COMs) are carbon-containing molecules with no less than six atoms \citep{Herbst2009}.
So far COMs have been detected in a wide range of ISM conditions, 
including cold cores and pre-stellar cores \citep{Bacmann2012, Cernicharo2012, Vastel2014, Jimenez-Serra2016},
low-mass star formation regions \citep{Bergner2017, Oberg2010, vanGelder2020},
high-mass star formation regions \citep{Li2017, Suzuki2018, Law2021},
protoplanetary disks \citep{Brunken2022},
comets \citep{Altwegg2016}, and carbonaceous meteorites \citep{Oba2022}.
Recently, the large number of complex molecules detected
in Taurus molecular cloud-1 \citep[see][and references therein]{Agundez2021} 
continues to increase the curiosity about their origin and formation in the early stages of star formation.

The formation mechanism of interstellar COMs has been an interesting question
for many astrochemists since these COMs were detected. 
\citet{Charnley1992} suggested that interstellar COMs were formed by their precursors in the gas phase.
These precursors, however, were formed on the dust grains and then sublimated as the temperature of grains increases. 
However, this scenario was challenged by new experimental and computational studies, 
which show that the gas-phase synthesis route was not efficient enough to produce the observed abundances of COMs,
such as methyl formate in hot cores and corinos \citep{Horn2004, Geppert2007}. 
\citet{Garrod2006} suggested that COMs were formed on dust grains and sublimated into gas phase as the temperature increases. 
In this approach, radicals are generated by photodissociation of icy species such as methoxy
and recombine with other radicals to form COMs when the dust temperature is between 20 K and 40 K. 
However, this COM formation scenario still underestimated the observed COMs abundances toward low-mass protostars \citep{Aikawa2008}. 
More recently, \citet{Lu2018} suggested a new COM formation scenario. 
In this new approach, radicals generated by photodissociation of icy species are frozen 
and stored in the ice mantle when the dust temperature is around 10 K.
With the increase of the temperature in subsequent stages, these frozen radicals recombine to form COMs. 
COM abundances toward low-mass protostars can be explained well by this approach \citep{Lu2018}.  

In recent years, the detection of COMs such as CH$_3$CHO and CH$_3$OCH$_3$ in the cold environments where 
the gas temperature $\sim$10 K suggested that COMs in the cold astronomical sources 
may be synthesized via a different mechanism because radicals cannot diffuse and
recombine to form COMs on dust grains at a temperature as low as 10 K \citep{Bacmann2012, Vastel2014, Agundez2021}.
On the other hand, the molecular complexity of comets may be related to the initial evolutionary 
stages of star formation \citep{Lefloch2018},
so there is increasingly strong interest in elucidating the mechanism of COMs formation in cold sources.
\citet{Vasyunin2013} and \citet{Balucani2015} 
suggested that the COMs detected in cold cores were formed in the gas phase,
with their precursors formed on grain surface and sublimated via reactive desorption \citep{Garrod2007}. 
One common problem with both COM formation scenarios is that 
the abundances of COM precursors such as CH$_3$O are overestimated \citep{Vasyunin2013, Balucani2015}. 
There are also a few COM formation mechanisms suggesting that 
COMs detected in cold cores are synthesized on dust grains by energetic processing.
\citet{Reboussin2014} considered the impulsive heating of dust grain by cosmic-rays, 
which could help radicals on grain surface diffuse and react with each other.
\citet{Shingledecker2018} added cosmic-ray-induced reactions to the chemical models, 
which could help COM formation on grains in cold cores.
There are also non-diffusive surface reaction mechanisms to explain COMs detected in cold cores.   
\citet{Ruaud2015} suggested that COMs in cold cores were formed by complex induced reaction and Eley-Rideal mechanisms 
while \citet{Chang2016} introduced a chain-reaction mechanism to form COMs on grains.

Recently, the distribution of dust grain size is considered in a few astrochemical models, which may provide 
an alternative solution to the question of COM formation in cold sources \citep{Acharyya2011,Pauly2016,Iqbal2018,Chen2018}.  
In these models, the surface chemistry of different sizes of dust grains has been studied. 
\citet{Pauly2016} found that the efficiency of CO$_2$ formation on smaller dust grains significantly 
increased because smaller grains are warmer, 
so it is easier for CO and OH to diffuse and recombine to form CO$_2$ on smaller grains.
\citet{Chen2018} (hereafter as paper \Rmnum{1}) recently studied the surface chemistry 
on stochastically heated grains in cold cores.
They found that small amounts of COMs can be formed on small dust grains with radii around 0.0069 $\mu$m 
because their temperature spikes are large enough so that radicals can diffuse on these grains.

To the best of our knowledge, surface chemistry on grains whose radii are less than 0.0069 $\mu$m 
has never been investigated in astrochemical models.
It was found that CO can sublime on grains with radii 0.0069 $\mu$m in the paper \Rmnum{1}. 
Surface species should sublime more quickly on grains smaller than 0.0069 $\mu$m because of their larger temperature fluctuations.
So far, the roles of these very small grains in the chemical evolution of molecular clouds are not clear. 
We do not know whether COMs can be efficiently produced on these grains or not.

The purpose of this work is two-fold. 
Firstly, compared with the paper \Rmnum{1}, which shows that COMs can be efficiently formed on the stochastically heated grains, we focus in this paper on how COMs formed on these grains can help to explain the observed gas-phase COMs under the cold core conditions and compare our predictions with observations.
We also extend the reaction network and distinguish COM isomers (such as HCOOCH$_3$ and CH$_2$OHCHO) 
for a more complete study of surface chemistry on stochastically heated grains.
The reactive desorption mechanism is included in models so that COMs formed on dust grains can sublime in cold cores.
Secondly, two different grain-size distributions are used to thoroughly investigate the effect of grain coagulation on the chemistry.
The grain-size distribution calculated by \citet{HY09}, which considers the coagulation of smaller grains, 
is used in our chemical models to investigate its effect on the clouds chemistry.
The size of the smallest grains has been extended to much smaller (down to 10$^{-3}$ $\mu$m) than that used in the paper \Rmnum{1}
so that we can study surface chemistry on these grains with significant temperature fluctuations.

The organization of this paper is the following. 
In Section \ref{sec:distrib}, we introduce the grain-size distribution used in this work 
while our chemical models are introduced in Section \ref{sec:models}. 
Section \ref{sec:heating} briefly introduces the heating and cooling of dust grains.
Our results are presented in Section \ref{sec:results}. 
Model results are compared with observations in Section \ref{sec:discussion}. 
Finally, discussions and conclusions are presented in Section \ref{sec:summary}.

\begin{figure*}
\resizebox{10cm}{8cm}{\includegraphics{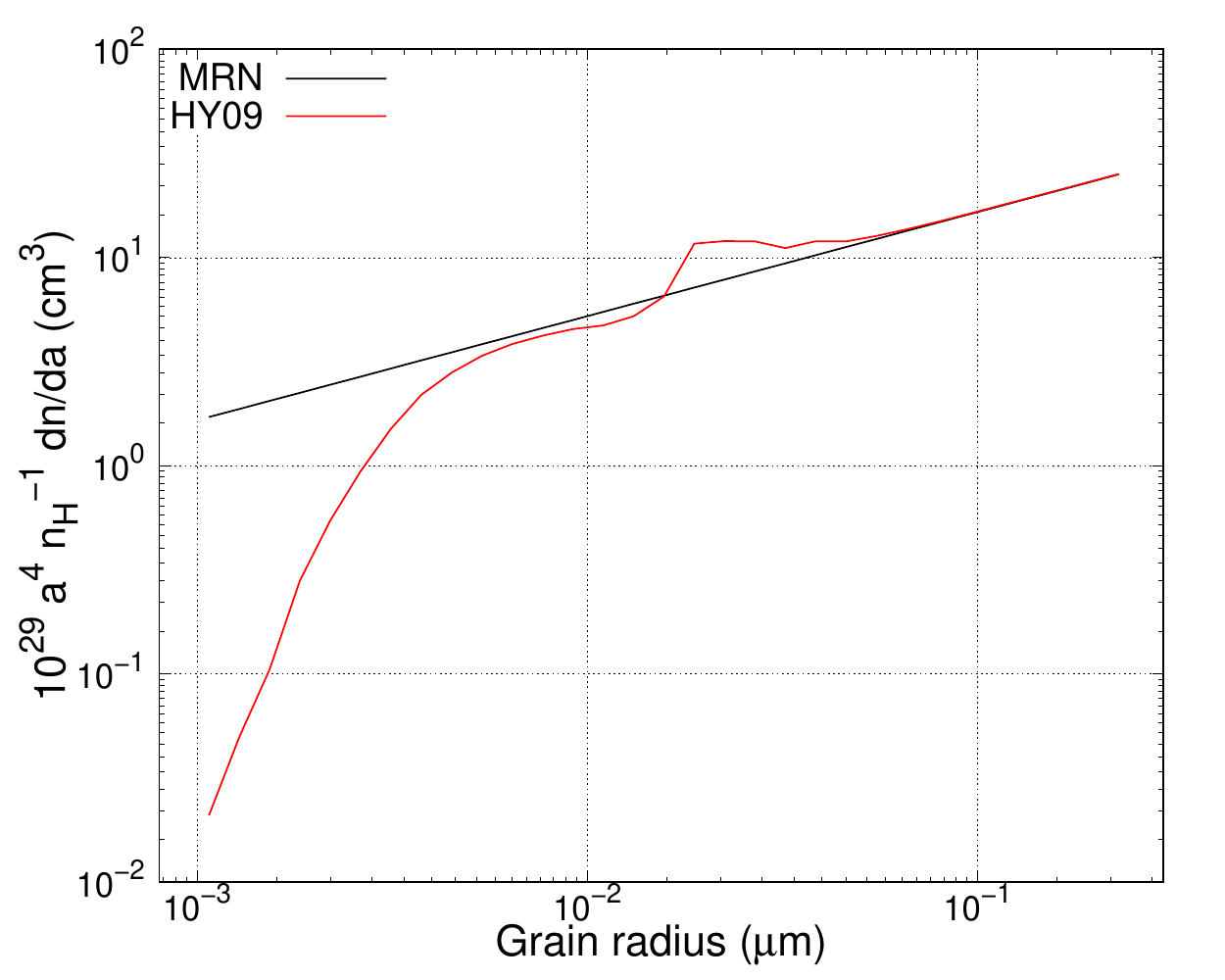}}
\caption{The MRN and HY09 grain-size distributions.
The HY09 grain-size distribution shown here corresponds to a dark cloud that has evolved 5 Myr \citep{HY09}.
}
\label{fig:sizeDistri}
\end{figure*}

\begin{table*}
\caption{The calculated MRN and HY09 grain-size distributions.}
\begin{tabular}{lllll}
\hline \hline
bin                    &   0                 & 1                  & 2                   & 3\\
\hline
MRN, radius ($\mu$m)   &   5.0314(-2)        & 1.3699(-2)         & 4.5404(-3)          & 1.5049(-3)\\
MRN, population        &   1                 & 15                 & 236                 & 3718\\
HY09, radius ($\mu$m)  &   4.7421(-2)        & 1.4404(-2)         & 4.7431(-3)          & 2.0928(-3)\\
HY09, population       &   1                 & 13                 & 139                 & 202\\
\hline
\label{tab:sizeDistri}
\end{tabular}
\medskip{\protect\\
Notes.\protect\\
a(b) means a $\times$ 10$^{b}$.}
\end{table*}

\section{Grain-size Distributions}\label{sec:distrib}
The distribution of dust grain size evolves with time due to the coagulation and shattering of grains in the
calculations by \citet{HY09}. However, it is not our purpose in this work to simulate dust coagulation and shattering 
concurrently with chemical evolutions. Therefore, following \citet{Ge2016}, we assume the distribution
of dust grain size varies little during the time period we run chemical simulations and adopt the calculated silicate 
grain-size distribution at the time 5 Myr in dark clouds (hereafter as HY09 distribution).
Fig.~\ref{fig:sizeDistri} shows the grain-size distribution used in this work. 
The range of dust grain radius is between $10^{-3}$ $\mu$m and $2.5\times10^{-1}$ $\mu$m.
The MRN grain-size distribution \citep{MRN} is also plotted in the same figure for comparison.
We can see that the population of small grains whose radii are less than 10$^{-2}$ $\mu$m in the MRN distribution 
is larger than that in the HY09 distribution due to the coagulation of small dust grains.

Similar to \citet{Pauly2016}, the HY09 and MRN distributions are divided into 5 bins, 
which are logarithmically equally spaced across the range of cross-section area.
The 0th bin has the biggest grains whose temperature fluctuations can be ignored, 
while the 4th bin has the smallest grains with the most significant temperature fluctuations.
The number of grains in the 0th bin is set to be 1. 
We calculate the number of grains in the 1st through the 4th bins and the representative grain radius
for each bin in the same way as we did in the paper \Rmnum{1}.  
The total number of grains in all bins are over 60000 for MRN distribution and over 6000 for HY09 distribution, 
so the computational cost to perform chemical simulations must be very expensive because of our limited CPUs.
However, the total number of grains can be reduced to minimize the computational cost. 
In the paper \Rmnum{1}, it was found that despite the temperature fluctuations induced by photons,
surface chemistry on grains whose radius are around 0.033 $\mu$m are large enough
so that surface chemistry on these grains is similar to that on the largest grains 
whose temperature fluctuations are ignored. 
The representative grain radius for the 1st bin is around 0.04 $\mu$m, 
so we can merge the 1st bin and the 0th bin into a new 0th bin.
Following the paper \Rmnum{1}, the number of grains in the new 0th bin is reset to 1, 
so the number of grains in other bins can be reduced. 
Tab.~\ref{tab:sizeDistri} shows the representative grain radius 
and population of grains in each bin for the HY09 and MRN grain-size distributions.
We can see that the population of smaller grains in the HY09 distribution is much less than
that in the MRN distribution. Moreover, other than the 0th bin, 
the representative grain radius for each bin in the HY09 distribution is slightly larger 
than that in the MRN distribution.
The grain surface provides sites for the adsorption of gas-phase species and the diffusion of grain-surface species.
Thus the surface area is important for the interactions between the gas-phase chemistry and the grain-surface chemistry.
From Tab.~\ref{tab:sizeDistri} we can calculate the grain surface area for each bin of grains.
The calculated fractional surface areas for the four bins relative to the total surface area for the MRN distribution are 14\%, 15\%, 26\%, and 45\%, respectively. And for the HY09 distribution, they are 25\%, 30\%, 35\%, and 10\%, respectively.
We can see that the fractional surface areas for the smallest grains in bin 3 of the MRN distribution occupy almost half of the total surface area. While for the HY09 distribution, they gradually increase in bins 0, 1, and 2, and occupy the majority of the fraction.

With the accretion of gas-phase species onto the dust grains, the radius of the grains increases with the accumulation of ice mantles. Thus, the grain surface area could also be changed.
The grain growth effect is considered in all our models, which has been explained in detail in the paper \Rmnum{1}.
In the following, two models (M1 and M2) are used to test the effect of different grain-size distribution on the chemistry.

\section{Chemical Models}\label{sec:models}
The physical conditions pertain to the cold cores where the terrestrial COMs were detected. 
The density of the hydrogen nuclei is n$_H$ = 1 $\times$ $10^5$ cm$^{-3}$,
the gas temperature is fixed to be 10 K, and the visual extinction is 10 mag. 
The cosmic ionization rate is $\zeta$ = 1.3 $\times$ 10$^{-17}$ s$^{-1}$. 
The initial elemental abundances are the same as used in the paper \Rmnum{1}, which are taken from \citet{Semenov2010}.

We use the macroscopic Monte Carlo method to perform the simulations, and the two-phase astrochemical model is adopted, including the gas-phase and grain-surface chemistry.
The chemical reaction network used in this work is based on that used in the previous paper \Rmnum{1}.
We modified the chemical reaction network as the following.
First, we distinguish methoxy (CH$_3$O) and hydroxymethyl (CH$_2$OH) radicals as in \citet{Chang2016}.
\citet{Cernicharo2012} reported the positive detection of CH$_3$O in the gas phase 
with a column density of $7\times10^{11}$ cm$^{-2}$ in the cold core B-1b,
and discussed its formation route, OH + CH$_3$OH $\rightarrow$ CH$_3$O + H$_2$O in the gas phase.
We include this gaseous formation route of CH$_3$O in the reaction network.
On the grain ice mantles, radicals CH$_3$O and CH$_2$OH are mainly 
produced by the photolysis of CH$_3$OH ice or the hydrogenation of H$_2$CO \citep{Garrod2008,Bennett2007,Bergantini2018}.
The photolysis of CH$_3$OH ice always produces CH$_3$O and CH$_2$OH at the same rate. 
Following \citet{Chang2016}, we also assume the hydrogenation of H$_2$CO on dust 
grains is able to produce CH$_3$O and CH$_2$OH at equal rate in one model (model M4), 
but in all other models, the hydrogenation of H$_2$CO can only produce CH$_2$OH.  
Secondly, the glycolaldehyde (CH$_2$OHCHO) reactions \citep{Garrod2015} are included in the reaction network. 
Finally, because CH$_3$OCH$_3$ is a newly added species, the destruction pathways for CH$_3$OCH$_3$ are also included
in the reaction network as in \citet{Chang2016}.

For the desorption mechanisms, thermal desorption, photo-desorption, and cosmic-ray desorption are considered.
The dust temperature is normally below COMs sublimation temperature in our models, 
so COMs formed on grains cannot enter the gas-phase by thermal desorption. 
In order to investigate how COMs formed on stochastically heated grains can help 
to explain the gas-phase COM abundances observed in the cold cores, 
the reactive desorption mechanism is considered in two models (models M3 and M4). 
Because the computational cost for each model is expensive 
while only a moderate amount of COMs are formed on stochastically 
heated grains in the previous paper \Rmnum{1}, the reactive desorption
efficiently is set to be 10\%, which is the highest efficiency used by \citet{Garrod2007}. 

In summary, the grain-size distribution used for models M1 and M2 is MRN and HY09, respectively. On the other hand, neither M1 nor M2 considers the reactive desorption. The probability for the hydrogenation of H$_2$CO to form CH$_3$O on the ice (the Y parameter) is 0 in both models. So we simulate models M3 and M4, which also use the HY09 grain-size distribution, but both models consider the reactive desorption, and the Y parameter used for M3 and M4 is set to 0 and 0.5, respectively.
Tab.~\ref{tab:models} summarizes all our models in this work.
We simulate models M1 and M2 one time. 
However, models M3 and M4 are simulated nine times with different random seeds 
to reduce the lowest fractional abundance down to 10$^{-12}$.

\begin{table*}
\caption{Chemical Models.}
\begin{tabular}{llll}
\hline \hline
Model  &   RD$^a$        & Y$^b$           & Grain-Size Distribution\\
\hline
M1     &   0             & 0               & MRN\\
M2     &   0             & 0               & HY09\\
M3     &   0.1           & 0               & HY09\\
M4     &   0.1           & 0.5             & HY09\\
\hline
\label{tab:models}
\end{tabular}
\medskip{\protect\\
Notes.\protect\\
$^a$Reactive desorption efficiency. \\
$^b$Probability for the hydrogenation of H$_2$CO to form CH$_3$O on the ice.}
\end{table*}

\begin{figure*}
\resizebox{16cm}{10cm}{\includegraphics{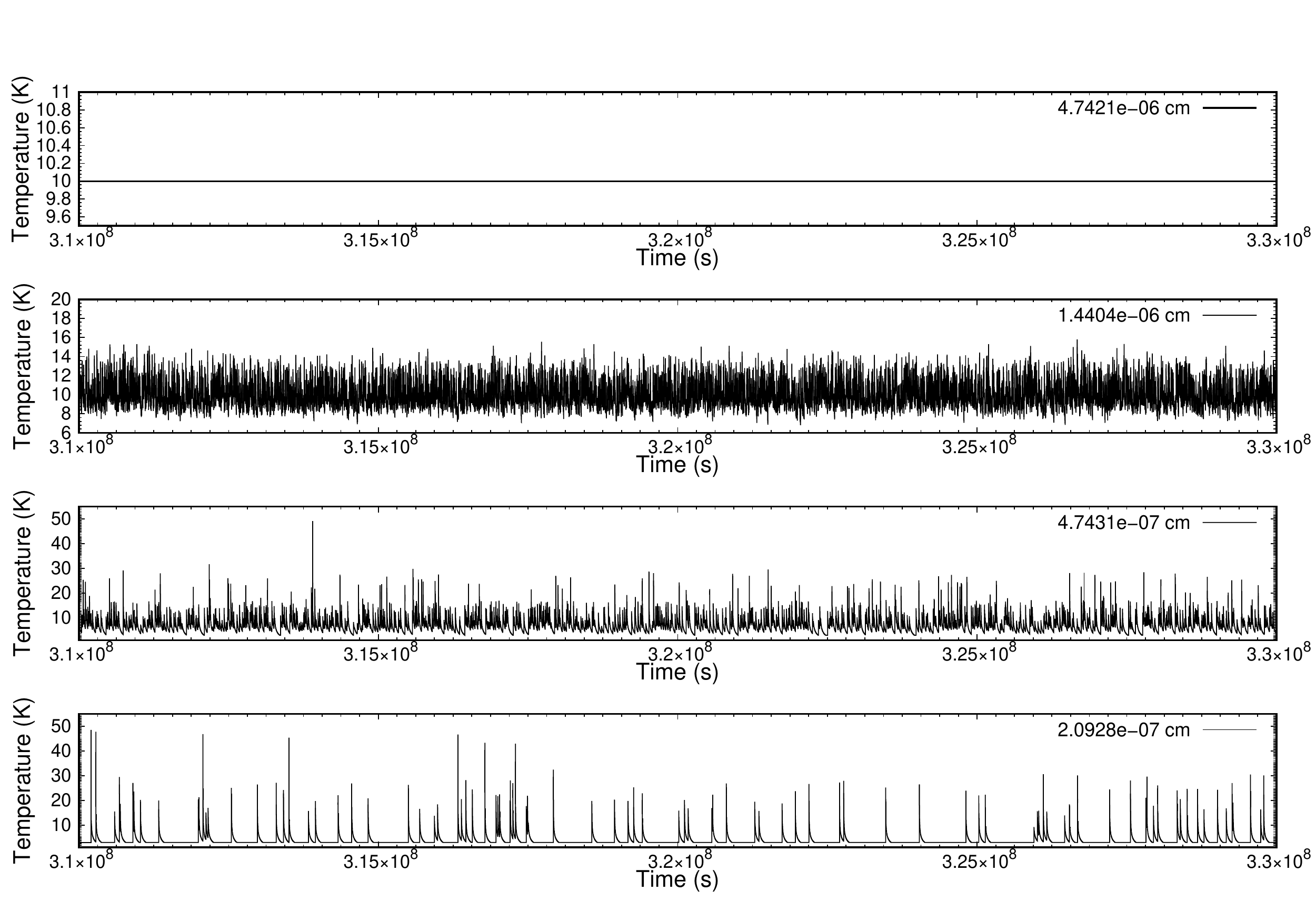}}
\caption{The grain temperature fluctuations for different sizes of grains in model M2.
The temperature of the biggest grain is fixed at 10 K. 
Whereas for other grains, they both include low energy photons heating and high energy photons heating, 
and the lowest grain temperature due to cooling is kept at CMB temperature of 3 K.
}
\label{fig:T_M2}
\end{figure*}

\section{Heating and Cooling of Dust Grains}\label{sec:heating}
Dust grains are assumed to be silicate and heated by both the low energy (infrared wavelength) 
and high energy (cosmic-ray induced secondary FUV) photons.
The wavelength of low energy photons are between 250 nm and 1 cm \citep{Cuppen2006}
while the spectrum of the high energy photons is complicated, so we set their wavelength to be the median
value between 850 \AA~ and 1750 \AA~ for a valid approximation (see paper \Rmnum{1}).
The rates of low energy photon heating events are calculated in the same way 
as that in the previous work \citep{Cuppen2006},
while the rates of high energy photon heating events are modified as the following.
A dust grain with radius $r$ is bombarded by high energy photons at a rate,
$Q_{abs}G_0F_0 \mathrm{\pi} r^2 $, where
$G_0 = 10^{-4}$ is a scaling factor for the high energy photons \citep{Shen2004},
$F_0 = 10^8 cm^{-2}s^{-1}$ is the standard interstellar radiation field and 
$Q_{abs}$ is the wavelength and grain-size dependent absorption coefficient \citep{Draine1984,Draine1985}.

We only consider radiative grain cooling, which is approximated by the continuous cooling in this work, 
because the cooling by CO sublimation is not likely to be important based on the estimation in paper \Rmnum{1}.

We follow \citet{Cuppen2006} to calculate the size-dependent heat capacity of grains. 
The calculation of the temperature of grains that have absorbed photons or undergo radiative cooling 
is straightforward \citep{Cuppen2006}.
Fig.~\ref{fig:T_M2} shows the calculated temperature fluctuations for different sizes of grains in model M2.
For grains in the same type (1st, 2nd or 3rd) of bin, 
the grain temperature fluctuations in model M1 are larger because of the smaller representative size of the grain.
Compared with the minimal grain used in paper \Rmnum{1} (5 $\times$ 10$^{-3}$ $\mu$m),
the minimal grain used in this work is smaller (down to 1 $\times$ 10$^{-3}$ $\mu$m),
which results in larger temperature fluctuations.
Due to the temperature fluctuations of different sizes of grains, the desorption rates of surface species on smaller grains are enhanced than that on the larger grains. Thus, the number of a gas-phase species in a cell containing a small grain may differ from the number of a gas-phase species in a cell containing a large grain. We further assume that the gas species are well mixed among the cells (i.e., the number of a gas-phase species in different cells is proportional to the volume of the cells).
Therefore, the mixing period, which is used to control the convergence of the gas-phase species 
among different types of cells, is also becoming shorter because of the more frequent interactions
between gas-phase species and grain-surface species (see \citet{Chen2018} for a detailed explanation).
A mixing period of 10 yr is used in this work, while a period of 1000 yr was used in previous work.

The macroscopic Monte Carlo approach is used to perform the simulations.
Following paper \Rmnum{1}, we use the parallel computation to accelerate simulations.
In this approach, each grain is put into a cell of gas. 
The initial total dust-to-gas mass ratio is fixed to 0.01 
while the initial population of gas-phase species in each cell is proportional 
to the surface area of each dust grain. 
We refer the paper \Rmnum{1} for details of our parallel computation approach. 
The multi-thread programming technique is also used to save CPUs as in the paper \Rmnum{1}.
In total, 228 CPUs are used to simulate model M1 while 38 CPUs are used to simulate other models. 
We used the Taurus High Performance Computing System of Xinjiang Astronomical Observatory 
to perform these simulations. It took about two weeks to simulate each chemical model.

\begin{figure*}
\resizebox{18cm}{14cm}{\includegraphics{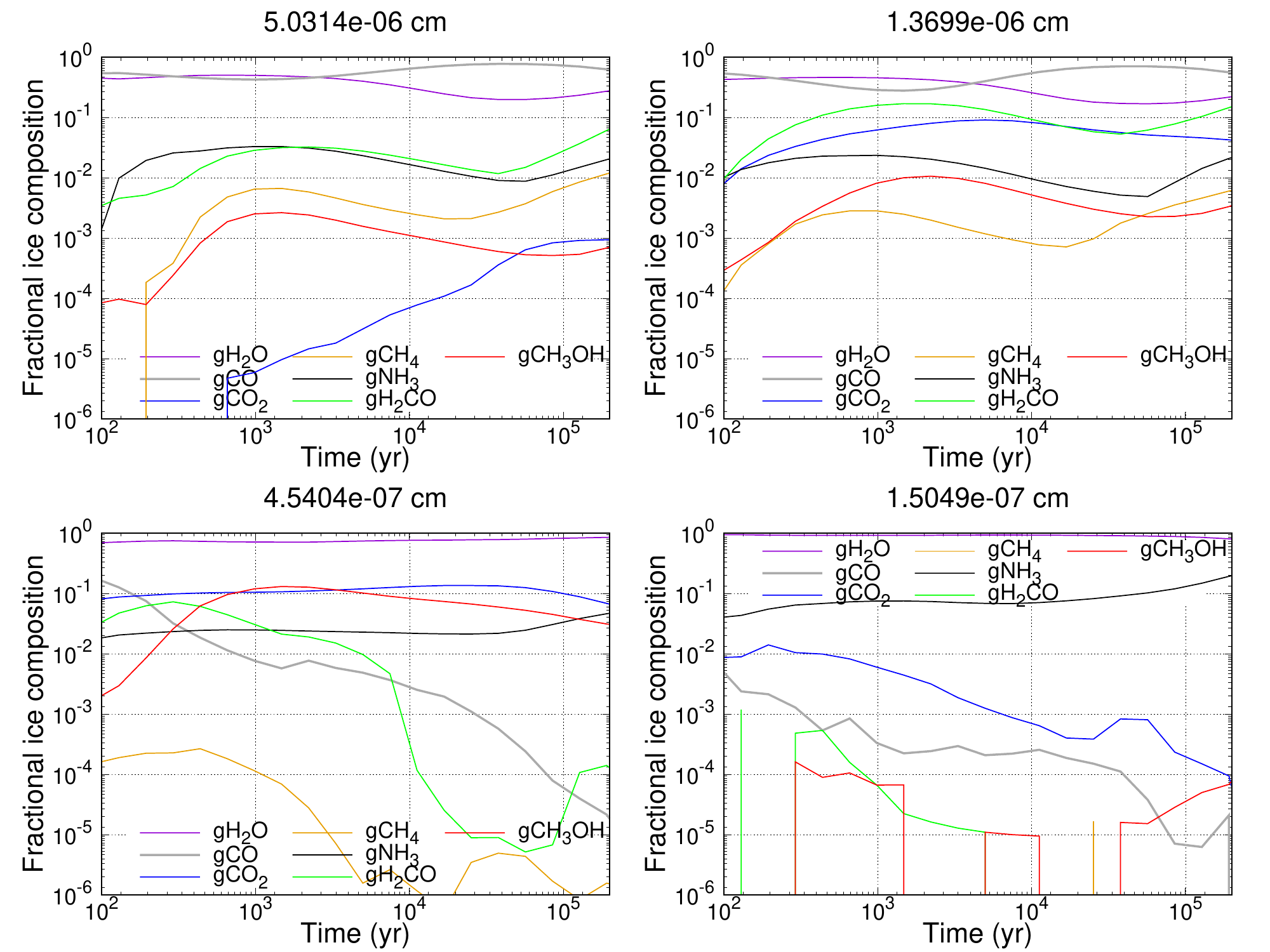}}
\caption{The temporal evolution of fractional ice composition on different sizes of grains in model M1.
In this figure and the following figures, ``g'' stands for granular species.
}
\label{fig:ice_M1}
\end{figure*}

\begin{figure*}
\resizebox{18cm}{14cm}{\includegraphics{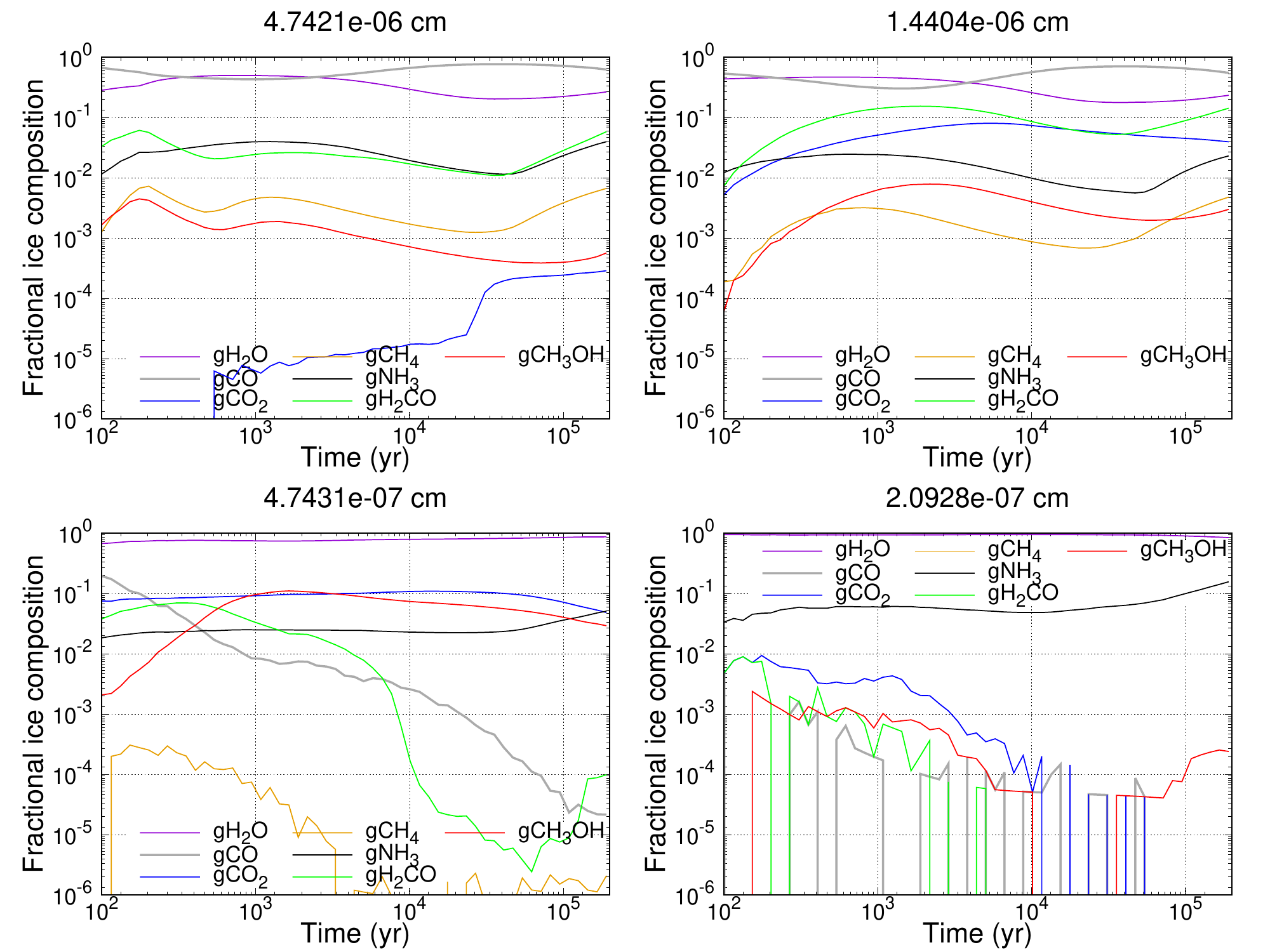}}
\caption{The temporal evolution of fractional ice composition on different sizes of grains in model M2.
}
\label{fig:ice_M2}
\end{figure*}

\begin{figure*}
\resizebox{15cm}{10cm}{\includegraphics{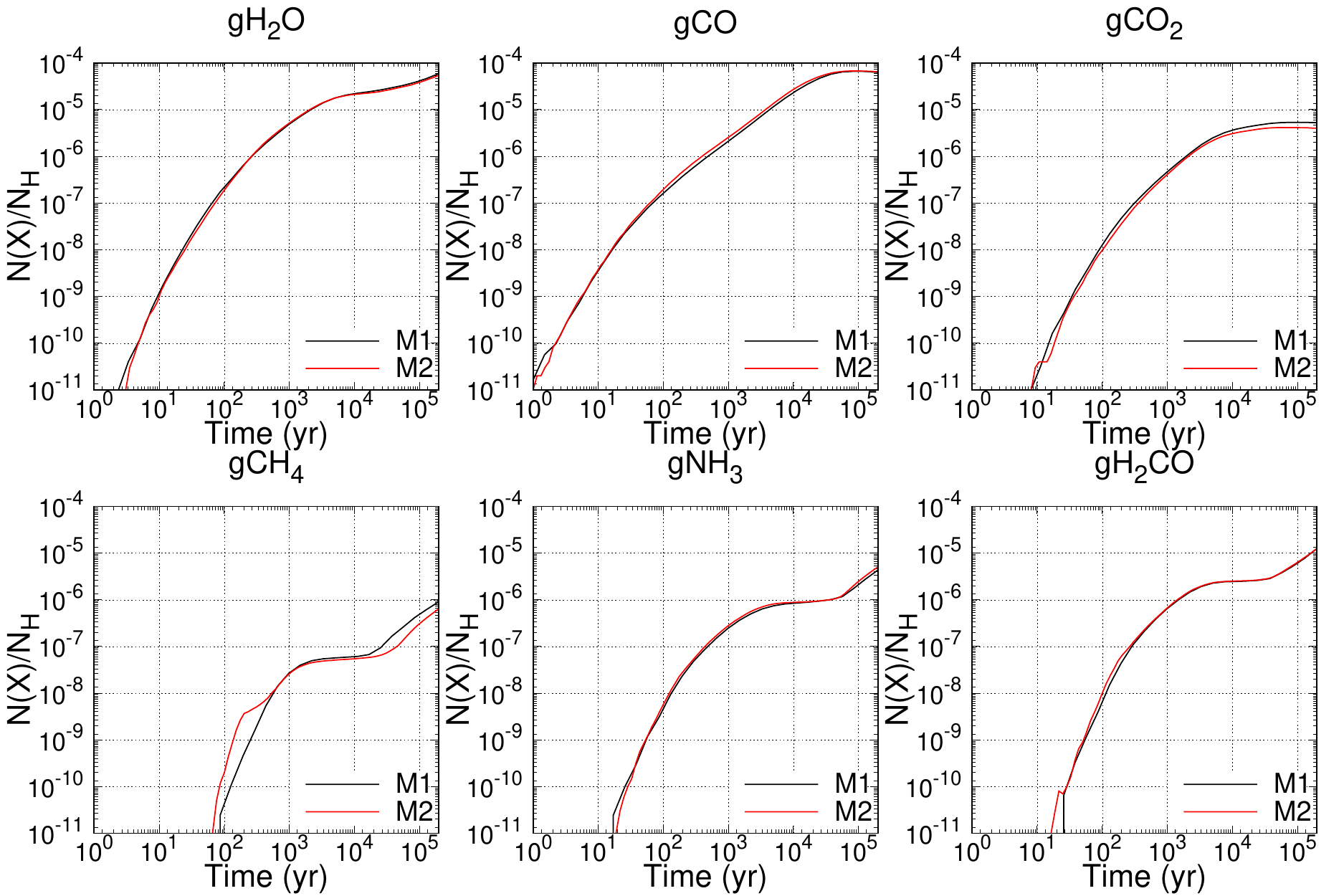}}
\caption{The total abundances of selected granular species as a function of time in models M1 and M2.
}
\label{fig:M1vsM2_grain}
\end{figure*}

\begin{figure*}
\resizebox{18cm}{14cm}{\includegraphics{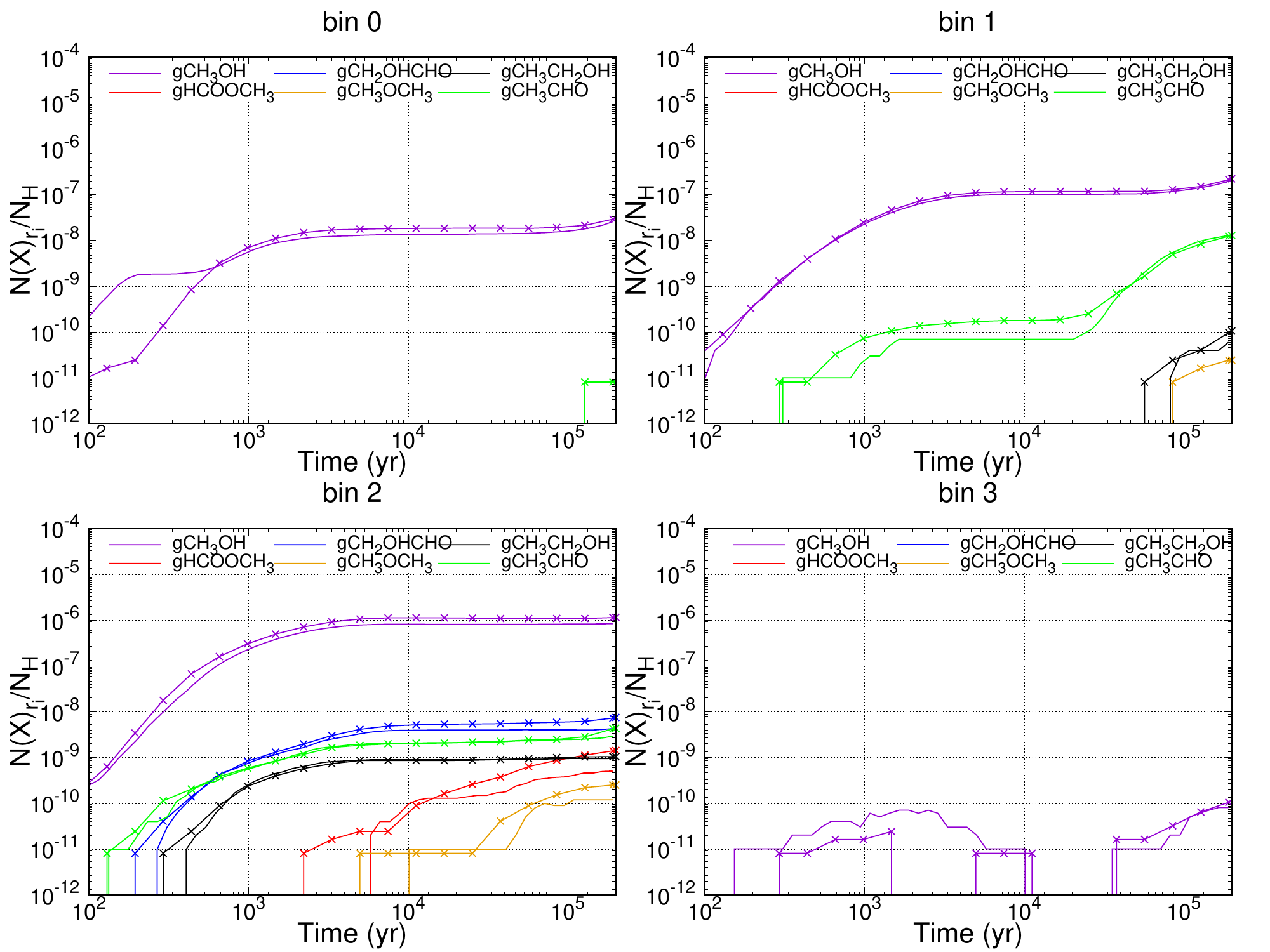}}
\caption{The fractional abundances of COMs as a function of time on different grains in models M1 and M2.
Lines with cross marks represent model M1, whereas those without cross marks are for model M2.
The grain size is represented by the bin. See Tab.~\ref{tab:sizeDistri} for details of 
the representative radii of grains in different bins.
}
\label{fig:diff_M1vsM2_gCOMs}
\end{figure*}

\begin{figure*}
\resizebox{15cm}{10cm}{\includegraphics{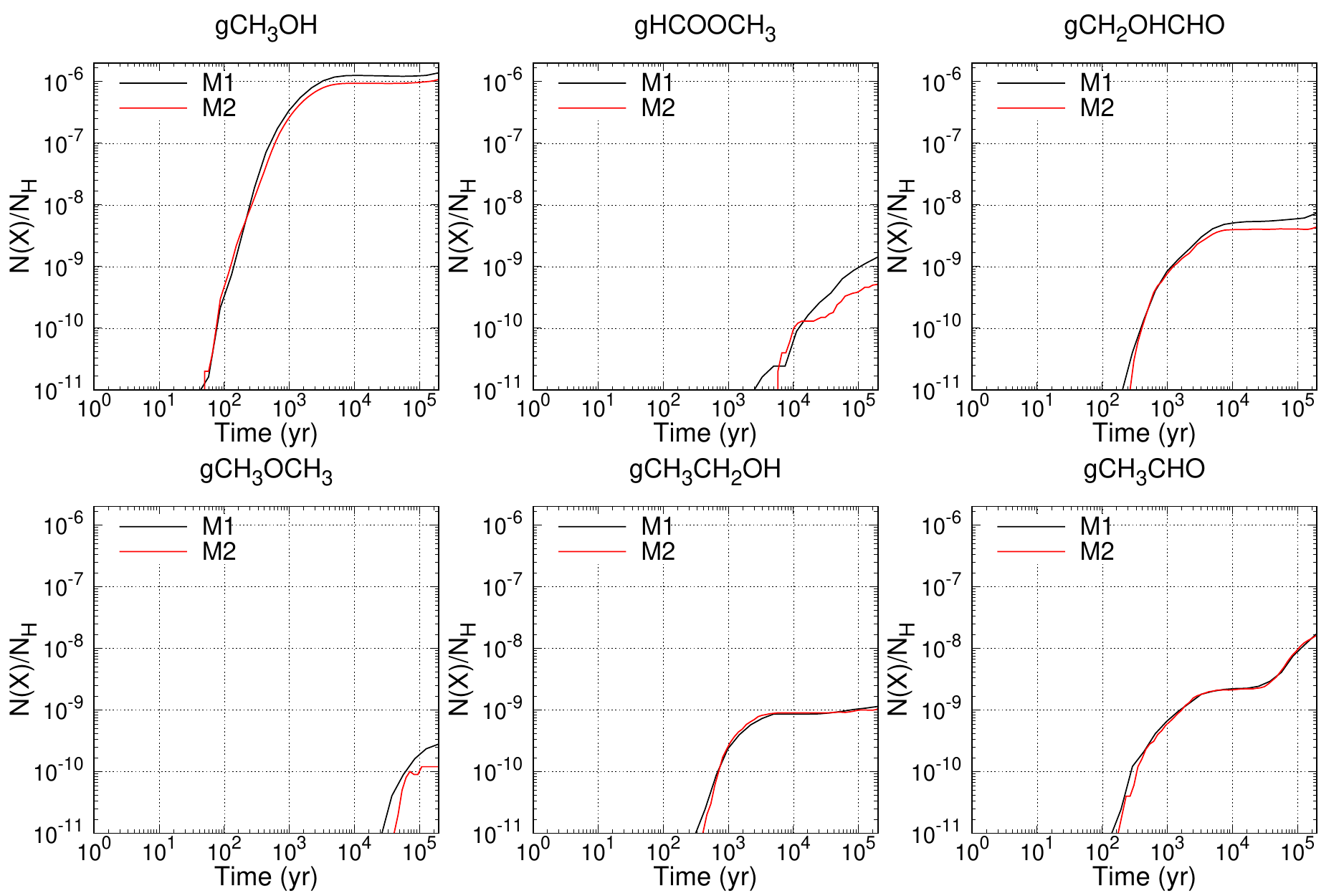}}
\caption{The total fractional abundances of granular COMs as a function of time in models M1 and M2.
}
\label{fig:M1vsM2_gCOMs}
\end{figure*}

\section{Results}\label{sec:results}

\subsection{The Effects of Grain Coagulation}
In this subsection, we study the effect of grain coagulation on the chemical evolution under the cold core conditions.  
We define two terms before reporting our simulation results. 
First, N(X)$_{r_i}$/N$_H$ represents the fractional abundance of a surface species X on grains with radius $r_i$, 
where N(X)$_{r_i}$ is the population of surface species X on all grains with radius $r_i$, 
and N$_H$ is the total population of H nuclei in the system. 
Secondly, the total fractional abundance of a species X is defined as, N(X)/N$_H$, 
where N(X) is the population of species X in all cells of gas. 
So for a surface species X, we have N(X)/N$_H$ = $\sum_{r_0}^{r_3} N(X)_{r_i}/N_H$.

\subsubsection{Major Ice Constituents}
The temporal evolution of fractional ice compositions on different sizes of grains in models M1 and M2
are shown in Fig.~\ref{fig:ice_M1} and Fig.~\ref{fig:ice_M2}, respectively.
Hereafter, we use the letter ``g'' to designate icy species. 
We only show the major ice constituents in these two figures.
The fractional ice composition of a granular species X on grains with radius $r_i$ 
is calculated as N(X)$_{r_i}$/N(major)$_{r_i}$ where N(X)$_{r_i}$ and N(major)$_{r_i}$ are the population of X
and all major granular species on all grains with radius $r_i$, respectively.

The fractional ice compositions on the grains other than the smallest one in model M1 are similar
to these in model M2 if the sizes of these grains are representative of the same type (0th, 1st, and 2nd) of bin.
Water ice is always the most abundant granular species on the two smallest grains at the time 2 $\times$ 10$^5$ yrs. 
Granular CO is one of the most abundant species on the two biggest sizes of grains in each grain-size distribution.
Due to the temperature fluctuations of the second biggest grains (i.e., grains in bin 1), 
surface species such as H, CO, and O diffuse faster on these grains 
than on the biggest grain (i.e., grain in bin 0) whose temperature is kept at 10 K.
As a result, it is easier to synthesize gCO$_2$, gH$_2$CO, and gCH$_3$OH on the second biggest grains.
As the grains become even smaller, the grain temperature fluctuations are large enough 
so that volatile surface species such as CO sublime quickly. 
So the fraction of gCO on the second smallest grains is lower than that on the two largest grains.

There are no more than one monolayer granular species on the smallest grains at any time in both models
because many surface species sublime quickly due to the large temperature fluctuations of these grains.
So it is difficult to synthesize gH$_2$CO, gCH$_3$OH, and gCH$_4$ on the smallest grains in both models.
Moderate amounts of gCO$_2$ can be synthesized on the smallest grains.
However, due to the grain temperature spike induced by the low and high energy photons can be as high as around 50 K 
and 90 K, respectively, gCO$_2$ formed on the smallest grains sublime quickly. 
So there are only a few gCO$_2$ molecules on the smallest grains at any time. 
At a later time, the populations of major surface species increase, 
but the population of gCO$_2$ molecules does not increase, so gCO$_2$ fraction decreases on the smallest grains at a later time.
The fluctuation of the fraction of gCO and gCO$_2$ on the smallest grains in the model M2 is 
much larger than that in model M1 because the population of the smallest grains 
in model M2 is much smaller than that in model M1. 
More gH$_2$CO and gCH$_3$OH molecules are formed on the smallest grains in model M2 than those in model M1 
because the smallest grains in model M2 are relatively cooler than those in model M1.
The major granular species on the smallest grains are gH$_2$O and gNH$_3$ in both models,
which are different comparing with those on bigger grains. 
Moreover, the fractions of these two major granular species on the smallest grains in model M1 are  
similar to those in model M2.

Fig.~\ref{fig:M1vsM2_grain} shows the total fractional abundances of selected major granular species 
as a function of time in models M1 and M2. 
The total fractional abundances of granular species in model M1 are similar to those in model M2. 
Because of the large temperature fluctuations of the smallest grains, few granular species reside on these grains. 
Thus, most granular species reside on bigger grains. 
Therefore, although there are differences in the fraction of granular species on the smallest grains, 
the total fractional abundances of granular species in models M1 and M2 are similar.

\subsubsection{Granular COMs}
Fig.~\ref{fig:diff_M1vsM2_gCOMs} shows the temporal evolution of the fractional abundances of selected COMs
on different sizes of grains in models M1 and M2. We use the bins instead of the grain radii to 
represent the sizes of grains in this figure (see our discussions in Section \ref{sec:distrib}.)
Bin 0 has the largest grains whose temperatures are kept at 10 K so that radicals can hardly diffuse, 
thus COMs other than methanol can hardly be produced on these grains in models M1 and M2. 
The temperature fluctuations of grains in bin 1 are large enough to produce significant amounts
of gCH$_3$CHO in both models. The difference of gCH$_3$CHO abundances in models M1 and M2 is not significant.
Moreover, at the time 2 $\times$ 10$^5$ yrs, the fractional abundances of gCH$_3$CHO on the 
second largest grains are almost the same in models M1 and M2.
The formation of COMs efficiently occurs on the second smallest grains in both models
because their temperature fluctuations are large enough so that radicals can diffuse on grains, 
but still small enough to keep most species on grains. 
We can see that the abundances of COMs other than gHCOOCH$_3$ on the second smallest
grains in models M1 and M2 are similar. 
Moreover, the difference in gHCOOCH$_3$ abundances in these two models is not significant. 
At the time 2 $\times$ 10$^5$ yrs, gHCOOCH$_3$ abundance in model M1 is a factor of 3 
larger than that in model M2 because the second smallest grains in model M1 are smaller,
thus the temperature fluctuations are larger, which helps radicals diffuse on grains. 
Few COMs are produced on the smallest grains in both models because these grains are overheated by photons. 
Therefore, grains in the bin 2 contribute the most to the production of granular COMs in models M1 and M2
because of their moderate grain temperature fluctuations.

Finally, Fig.~\ref{fig:M1vsM2_gCOMs} shows the total 
fractional abundances of selected granular COMs as a function of time. 
We can see that the total fractional abundances of COMs in models M1 and M2 are similar. 
Therefore, we can conclude that the coagulation of dust grains 
can hardly change COMs formation on dust grains.

\subsubsection{Gas-Phase Simple Species}
\begin{figure*}
\resizebox{15cm}{10cm}{\includegraphics{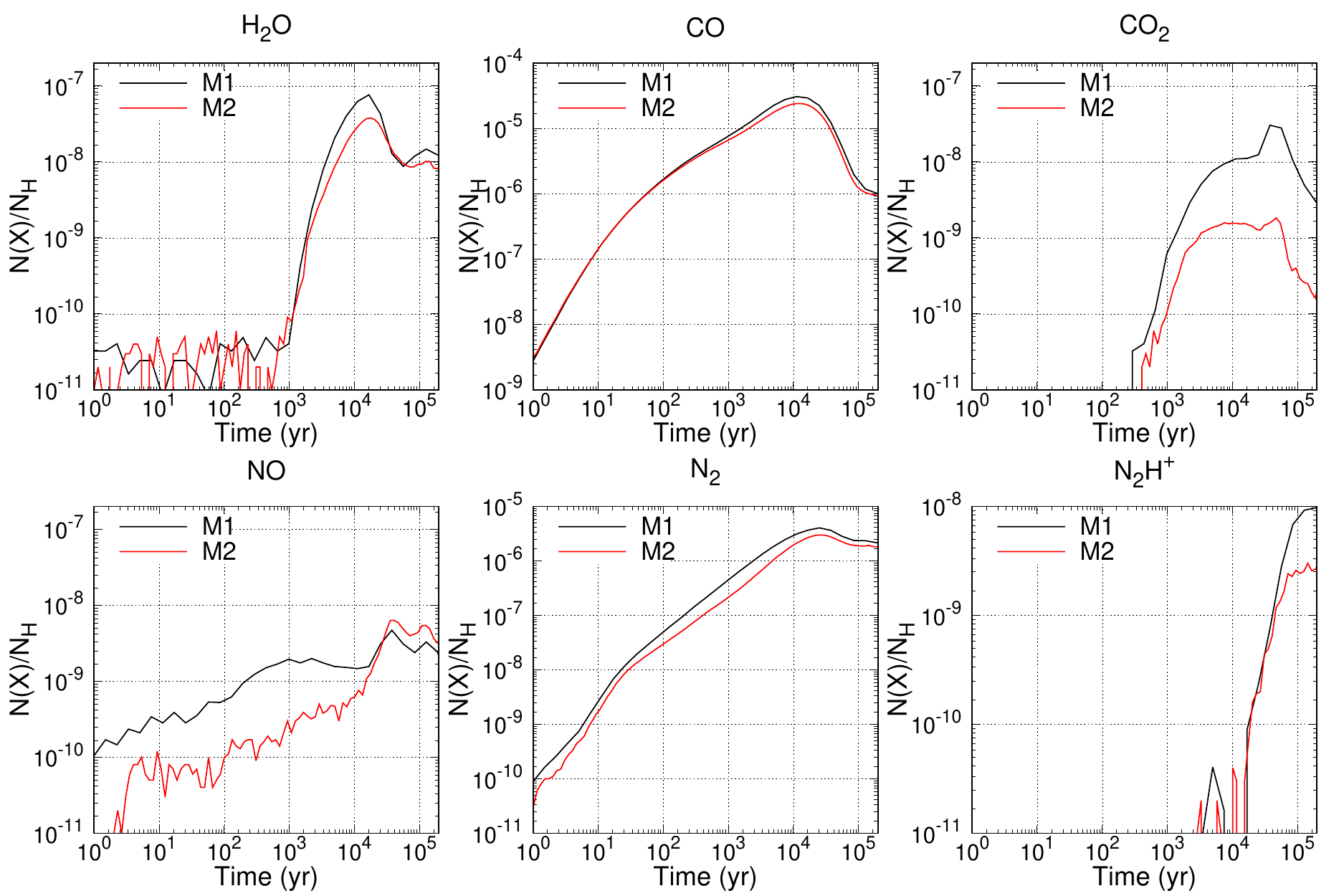}}
\caption{The total fractional abundances of selected gas-phase species as a function of time in models M1 and M2.
}
\label{fig:M1vsM2_gas}
\end{figure*}

The effect of grain coagulation can affect the gas-phase simple species in cold cores as shown in Fig.~\ref{fig:M1vsM2_gas},
which shows the total fractional abundances of selected gas-phase species as a function of time in models M1 and M2. 
The total fractional abundances of H$_2$O, CO, and N$_2$ in model M1 are similar to those in model M2.
However, the gaseous CO$_2$ and N$_2$H$^+$ abundances in models M1 and M2 can differ much.
The difference in the total fractional abundance of NO in models M1 and M2 is not significant after the time 2 $\times$ 10$^4$ yrs
although the difference can be a factor of a few before that time.
The difference of gaseous CO$_2$ abundance in models M1 and M2 can be as larger as one order of magnitude, 
which occurs around the time 4 $\times$ 10$^4$ yrs. The difference can be explained as the following.
Carbon dioxide can be efficiently formed on the smallest grains in models M1 and M2
while gCO$_2$ formed on the grain surface may sublime efficiently.
On the other hand, the population of grains in bin 3 of M1 is over one order of magnitude larger than that in bin 3 of M2,
so more gaseous CO$_2$ molecules are produced in model M1 than those in model M2.
The difference of N$_2$H$^+$ abundance in models M1 and M2 is about a factor of four after 1 $\times$ 10$^5$ yrs.
N$_2$H$^+$ originates from the gas phase and is mainly produced by the destruction of N$_2$ by H$_{2}^+$ and H$_{3}^+$.

Although grain coagulation should decrease the depletion rates of gas-phase species 
due to the reduction of the total grain surface areas, 
the abundance of CO in model M2 is only slightly lower than that in model M1. 
The controversy can be explained as the following.
Although the coagulation of the smallest grains does reduce the total grain surface areas, 
the smallest grains in models M1 and M2 are ``hot'' enough 
so that few species can reside on them for a significantly long time. 
Therefore, the smallest grains play little role in the depletion of volatile species such as CO and N$_2$. 
Moreover, volatile species such as gCO sublimate more quickly 
on the second smallest grains in model M1 because of the larger temperature fluctuations than that in model M2.  
Therefore, CO molecules in model M2 deplete relatively more quickly than those in model M1.

In summary, the effect of grain coagulation can affect some gas-phase species
by the number of small grains and their temperature fluctuations.
The most affected gas-phase species have the following characteristics: 
1) the contribution of their gas-phase production in the cells of large grains is not significant,
i.e~their abundances are relatively low at low gas-phase temperatures;
2) they are derived by the gas-phase chemistry in the cells of small grains, while the precursors are volatile 
and originate from the surface of small grains.

\subsection{Gas-Phase and Granular COMs in Models M3 and M4}
\begin{figure*}
\resizebox{18cm}{14cm}{\includegraphics{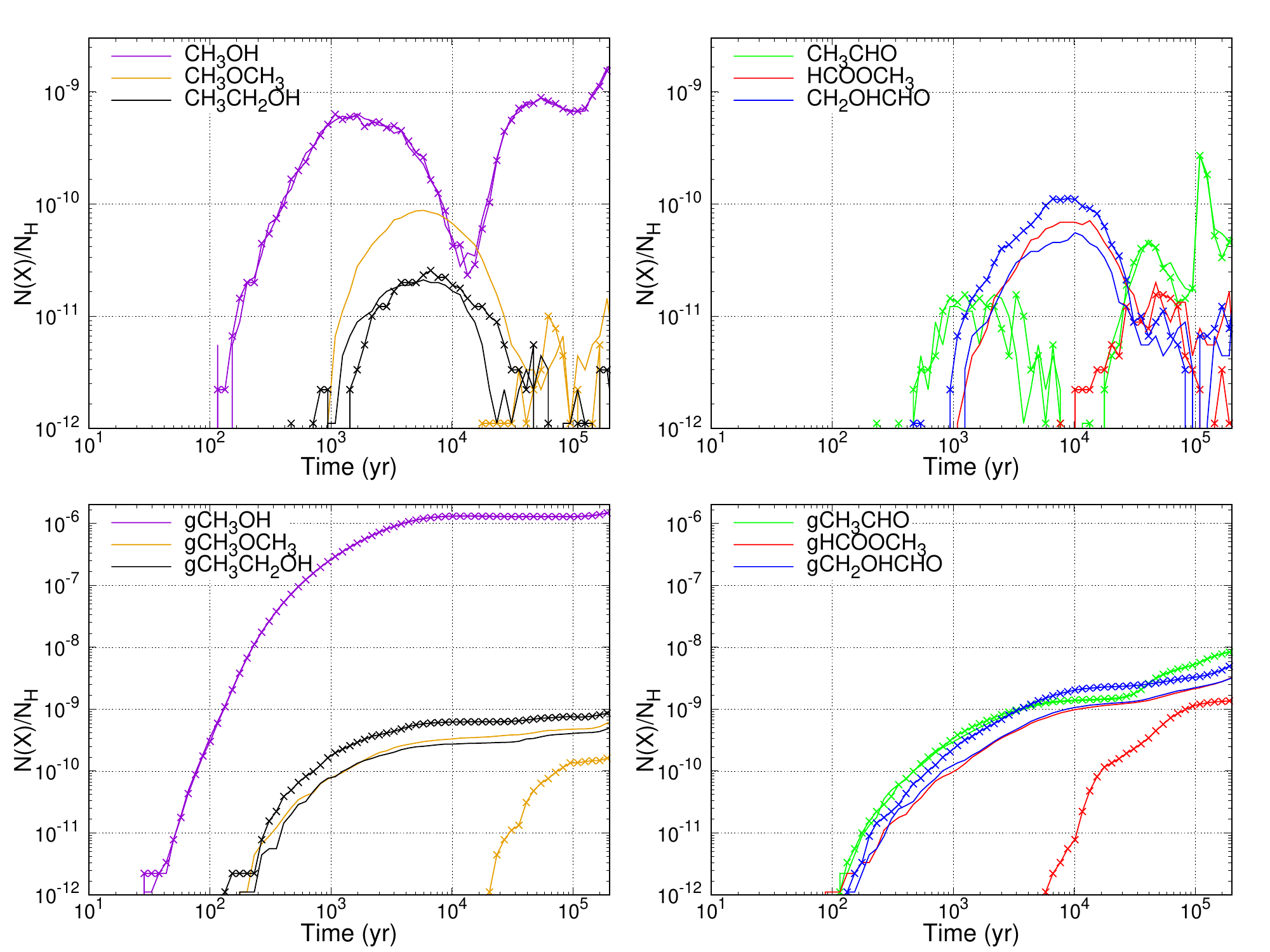}}
\caption{The COMs abundances in gas phase (upper panels) and in grain ice mantles (lower panels) for models M3 and M4.
Lines with cross mark style are for model M3, whereas those with no mark style are for model M4.
}
\label{fig:M3vsM4_COMs}
\end{figure*}

Fig.~\ref{fig:M3vsM4_COMs} shows the temporal evolution of selected gas-phase and granular COMs fractional abundances 
in models M3 and M4, which consider the reactive desorption mechanism. 
The efficiency of reactive desorption is fixed to 0.1.
The radical gCH$_3$O can be more efficiently synthesized in model M4 than that in model M3 
while the formation of the radical gCH$_2$OH is more efficient in model M3.
Therefore, more gHCOOCH$_3$ and gCH$_3$OCH$_3$ molecules are produced in model M4 
while the abundances of gCH$_2$OHCHO and gCH$_3$CH$_2$OH in model M3 are higher than those in model M4.
The total fractional abundances of other granular COMs in models M3 and M4 
are similar because the formation of these COMs is not dependent on gCH$_2$OH or gCH$_3$O.
Granular gCH$_3$OH abundance increases quickly before the time of 10$^4$ yr, 
but after that, it only slightly increases in both models.
On the other hand, the total fractional abundances of gHCOOCH$_3$ and gCH$_3$OCH$_3$ always increase
before the time at 2 $\times$ 10$^5$ yrs.

The abundances of the selected gas-phase COMs abundances vary a lot. 
Other than the time around 10$^4$ yrs, CH$_3$OH is the most abundant COM and its fractional abundance can be more than 10$^{-9}$.
CH$_3$CHO can be formed on grain surface and then comes to the gas phase via the reactive desorption.
On the other hand, CH$_3$CHO can also be efficiently synthesized in the gas phase by O + C$_2$H$_5$ $\rightarrow$ CH$_3$CHO + H.
So, after the time 2 $\times$ 10$^4$ yrs, CH$_3$CHO is more abundant than almost all COMs other than CH$_3$OH. 
Other than CH$_3$CHO and CH$_3$OH, the selected gas-phase COMs are mainly formed on the moderately heated grains, 
i.e., grains in bin 2, via the radical-radical recombination reactions. 
Because the total surface area of grains in bin 2 only accounts for 35\% of the overall grain surface area for model M4,
these COM fractional abundances are typically low (around 10$^{-11}$).

\section{Comparison with Observations}~\label{sec:discussion}
Many COMs have been detected in the cold dense sources \citep{Agundez2021,Jimenez-Serra2021}. The presence of COMs in these environments with typical low gas temperatures and high-density physical conditions hints that new COMs formation mechanism do not rely on the radical-radical recombination on moderately warm (20--40 K) grains whose radii are about 0.1 $\mu$m should be introduced in models. Many previous studies focused on the formation of COMs on the large grains, while the effect of small grains was less considered. With the help of the temperature fluctuations of small grains, the formation efficiency of COMs on the grain surface could be increased, which may further enrich the COMs observed in the gas phase.
One of the primary purposes of this work is to investigate how the COMs formed on the stochastically heated small grains 
may help to explain the observed COMs in these cold dense environments.

\begin{figure*}
\resizebox{18cm}{12cm}{\includegraphics{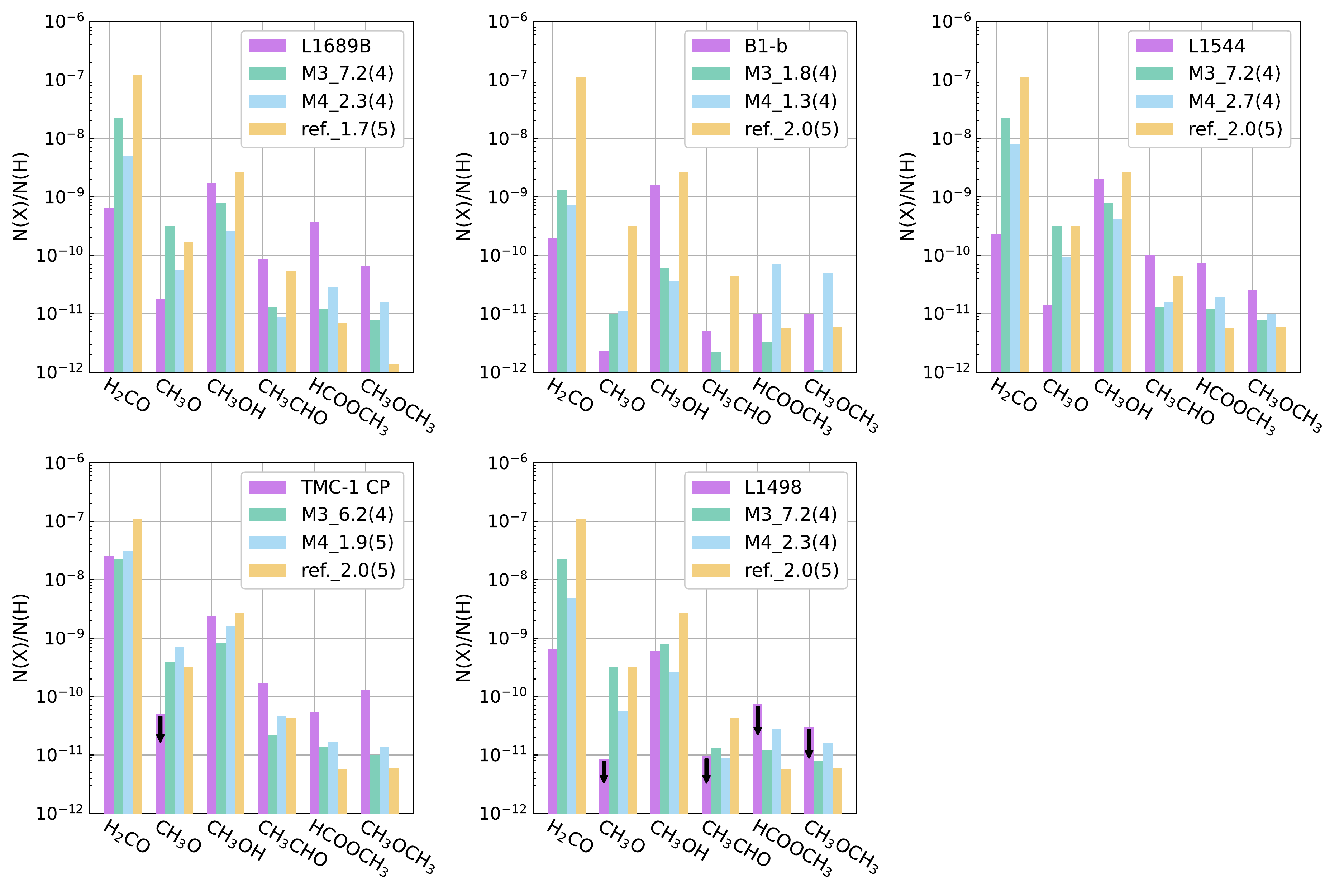}}
\caption{The comparison of different gaseous molecules in models M3, M4 and the reference model
with the observational data in the five sources, L1689B, B1-b, L1544, TMC-1 CP and L1498.
The legend in the figure after each model name represents the best fitting time, where a(b) = a $\times$ 10$^b$.
The black arrows represent the upper limits for some observed molecules.
}
\label{fig:erfc}
\end{figure*}

We select five cold dense sources (i.e., L1689B, B1-b, L1544, TMC-1 CP, and L1498) with typical low temperature and high gas density for our comparison between the model predictions and observations.
Tab.~\ref{tab:erfc} shows the observed gas-phase molecular abundances for formaldehyde, methoxy, and four COMs (CH$_3$OH, CH$_3$CHO, HCOOCH$_3$, and CH$_3$OCH$_3$), as well as our predicted model abundances at the best agreement time.
Fig.~\ref{fig:erfc} shows these data in a more vividly version.
Other than the models studied in this work, 
we also include the COM abundances predicted by a reference model that adopts a single type of standard grain with a radius of 0.1 $\mu$m commonly used in the astrochemical simulations.
The temperature of the grain is fixed to 10 K, and the Y parameter and the reactive desorption efficiency is 0 and 0.1, respectively.
The chemical network and other physical and chemical parameters used in the reference model are the same as the descriptions in Section \ref{sec:models}.
In the reference model, the number of H nuclei is about 6 $\times$ 10$^{11}$, 
so the lowest fractional abundance with respect to H nuclei is 1.7 $\times$ 10$^{-12}$,
which is comparable with the smallest observational value in the Tab.~\ref{tab:erfc}.
We run the reference model six times using different random seeds to smooth the evolutionary trends
for the COM abundances in order to compare with the observational data.
As for the model of M3 and M4, the total number of H nuclei is about 1 $\times$ 10$^{11}$,
so the lowest fractional abundance with respect to H nuclei is 10$^{-11}$.
So we run these two models nine times with different random seeds 
to reduce the lowest fractional abundance down to $\sim$10$^{-12}$.
The time of best agreement is calculated by using the complementary error function 
analysis \citep{Garrod2007, Chang2016}, which is expressed by the following formula,
\begin{equation}
k_{i}(t) = erfc(\frac{| log(X_{i}^{mod}(t)) - log(X_{i}^{obs}) |}{\sqrt{2} \sigma}),
\end{equation}
where $X_{i}^{mod}(t)$ and $X_{i}^{obs}$ are the abundance predicted by a model at time $t$ 
and the observed abundance for specie $i$, respectively. 
$\sigma$ is the standard deviation and represents the uncertainty of the observed abundance. Its value is set to 1, corresponding to one order of magnitude higher or lower between the observed and predicted abundances.
The largest value of $k_{i}$ is 1 when the predicted abundance and the observed abundance of a species are the same.
Thus, we could determine the best agreement time for our model.

We can see in Tab.~\ref{tab:erfc} that
the chemical timescales of the best agreement for models M3 and M4 are pretty early.
On the one hand, the evolutionary time of the sources is not well determined.
TMC-1 CP is one of the most intensively studied sources, and its chemical timescale is $\sim$5 $\times$ 10$^5$ yrs 
as indicated by observations and chemical models \citep{Pineda2010,Agundez2013,Chen2022}.
L1544 is an early-stage prestellar core, and its typical timescale could range from $\sim$10$^4$ to 1--3 $\times$ 10$^5$ yrs 
as constrained by CO depletion \citep{Caselli1999}, 
the freeze-out of NH$_2$D \citep{Caselli2022,Spezzano2022} and deuteration molecules \citep{Spezzano2022}.
Thus, we can roughly conclude that L1544 is younger than TMC-1 CP.
It can be noticed in Tab.~\ref{tab:erfc} that 
the predicted chemical timescale for TMC-1 CP indeed approaches the value in the literature,
and the predicted value for L1544 is smaller than that of TMC-1 CP, 
although the predicted timescale for L1544 tends to be lower than the values in the literature.
Note that our physical parameters are not designed for specific sources, 
but pertain to the general properties of the cores.
Therefore, the interpretation of the results should also be general.
Overall, model M4 performs better to fit the observed molecular abundances 
toward all sources than model M3 and the reference model do.

The differences between the observed CH$_3$OCH$_3$ abundances toward all sources 
and that predicted by model M4 are within one order of magnitude.
Although model M3 can reproduce CH$_3$OCH$_3$ abundances toward most sources in the table,
its prediction is one order of magnitude lower than that of TMC-1 CP.
Both models could well reproduce the observed abundances of HCOOCH$_3$ toward almost all sources except for L1689B, 
and the observed abundances of CH$_3$OH toward almost all sources except for B1-b.
The abundances of CH$_3$CHO are well reproduced by both models.

Both H$_2$CO and CH$_3$O are the precursors of the COMs shown in the table except for CH$_3$CHO.
Instead of underestimating the selected COMs abundances, 
the abundances of these two precursors are overestimated by models M3 and M4.
Model M3 performs worse than M4 does. 
Model M3 overestimates the abundances of these two precursors toward three sources (L1689B, L1544, and L1498).
While model M4 can reproduce the abundances of these two precursors toward most of the sources
except that the abundance of CH$_3$O in TMC-1 CP is overestimated.

Finally, the reference model results suggest that
the best fitting time is around $\sim$10$^5$ yr for the five sources.
The reference model also overestimates the abundances of H$_2$CO and CH$_3$O toward almost all sources.
At the time of best fit, the abundances of H$_2$CO and CH$_3$O predicted by the reference model are between 
a factor of a few and more than two orders of magnitude higher than these by models M3 and M4.
On the other hand, the abundances of HCOOCH$_3$ and CH$_3$OCH$_3$ are underestimated by the reference model.
The abundances of these two COMs predicted by models M3 and M4 at the best fitting time
are typically a factor of few larger than that by the reference model.
It should be noted that the formation pathway for gaseous HCOOCH$_3$ and CH$_3$OCH$_3$ 
in the reference model is different comparing with that in models M3 and M4.
In models M3 and M4, HCOOCH$_3$ and CH$_3$OCH$_3$  
are formed on the stochastically heated grains and then sublimate via reactive desorption.
In the reference model, however, the gaseous HCOOCH$_3$ and CH$_3$OCH$_3$ are produced directly in the gas phase.
Methanol can be efficiently formed on cold ($\sim$10 K) dust grain surfaces, therefore, 
the reference model performs even better than models M3 and M4 to reproduce the observed CH$_3$OH abundances 
toward all sources in the table. 
Moreover, because CH$_3$CHO can be efficiently synthesized in the gas phase, 
models M3 and M4 do not perform better than the reference model to reproduce its abundance toward these sources.

\begin{table*}
\caption{Comparison between the model predictions and the observed abundances 
for formaldehyde, methoxy and four COMs from five selected cold dense sources.}
\begin{tabular}{lllllllll}
\hline \hline
source     & obs./mod. & Time (yr) & H$_2$CO        & CH$_3$O         & CH$_3$OH        & CH$_3$CHO     & HCOOCH$_3$      & CH$_3$OCH$_3$ \\
\hline      
L1689B     & obs.      & -         & 6.5(-10)$^a$   & 1.8(-11)$^b$    & 1.7(-9)$^b$     & 8.5(-11)$^a$  & 3.7(-10)$^a$    & 6.5(-11)$^a$ \\
           & M3        & 7.2(4)    & {\bf 2.2(-8)}  & {\bf 3.2(-10)}  & 7.8(-10)        & 1.3(-11)      & {\bf 1.2(-11)}  & 7.8(-12)\\
           & M4        & 2.3(4)    & 4.9(-9)        & 5.7(-11)        & 2.6(-10)        & 8.9(-12)      & {\bf 2.8(-11)}  & 1.6(-11) \\
           & ref.      & 1.7(5)    & {\bf 1.2(-7)}  & 1.7(-10)        & 2.7(-9)         & 5.4(-11)      & {\bf 7.0(-12)}  & {\bf 1.4(-12)} \\
\hline      
B1-b$^a$   & obs.      & -         & 2.0(-10)       & 2.3(-12)        & 1.6(-9)         & 5.0(-12)      & 1.0(-11)        & 1.0(-11) \\
           & M3        & 1.8(4)    & 1.3(-9)        & 1.0(-11)        & {\bf 6.0(-11)}  & 2.2(-12)      & 3.3(-12)        & 1.1(-12)\\
           & M4        & 1.3(4)    & 7.3(-10)       & 1.1(-11)        & {\bf 3.7(-11)}  & 1.1(-12)      & 7.1(-11)        & 5.0(-11) \\
           & ref.      & 2.0(5)    & {\bf 1.1(-7)}  & {\bf 3.2(-10)}  & 2.7(-9)         & 4.4(-11)      & 5.7(-12)        & 6.0(-12) \\
\hline      
L1544      & obs.      & -         & 2.3(-10)$^c$   & 1.4(-11)$^d$    & 2.0(-9)$^c$     & 1.0(-10)$^d$  & 7.5(-11)$^e$    & 2.5(-11)$^d$ \\
           & M3        & 7.2(4)    & {\bf 2.2(-8)}  & {\bf 3.2(-10)}  & 7.8(-10)        & 1.3(-11)      & 1.2(-11)        & 7.8(-12) \\
           & M4        & 2.7(4)    & 7.8(-9)        & 9.4(-11)        & 4.2(-10)        & 1.6(-11)      & 1.9(-11)        & 1.0(-11) \\
           & ref.      & 2.0(5)    & {\bf 1.1(-7)}  & {\bf 3.2(-10)}  & 2.7(-9)         & 4.4(-11)      & {\bf 5.7(-12)}  & 6.0(-12) \\
\hline
TMC-1 CP   & obs.      & -         & 2.5(-08)$^f$   & $<$5.0(-11)$^f$ & 2.4(-09)$^g$    & 1.7(-10)$^g$  & 5.5(-11)$^h$    & 1.3(-10)$^h$ \\
           & M3        & 6.2(4)    & 2.2(-08)       & 3.9(-10)        & 8.3(-10)        & 2.2(-11)      & 1.4(-11)        & {\bf 1.0(-11)} \\
           & M4        & 1.9(5)    & 3.1(-08)       & {\bf 7.0(-10)}  & 1.6(-09)        & 4.7(-11)      & 1.7(-11)        & 1.4(-11) \\
           & ref.      & 2.0(5)    & 1.1(-07)       & 3.2(-10)        & 2.7(-09)        & 4.4(-11)      & 5.7(-12)        & {\bf 6.0(-12)} \\
\hline
L1498      & obs.      & -         & 6.5(-10)$^i$   & $<$8.5(-12)$^j$ & 6.0(-10)$^j$    & $<$9.5(-12)$^j$  & $<$7.5(-11)$^j$    & $<$3.0(-11)$^j$ \\
           & M3        & 7.2(4)    & {\bf 2.2(-08)} & {\bf 3.2(-10)}  & 7.8(-10)        & 1.3(-11)      & 1.2(-11)        & 7.8(-12) \\
           & M4        & 2.3(4)    & 4.9(-09)       & 5.7(-11)        & 2.6(-10)        & 8.9(-12)      & 2.8(-11)        & 1.6(-11) \\
           & ref.      & 2.0(5)    & {\bf 1.1(-07)} & {\bf 3.2(-10)}  & 2.7(-09)        & 4.4(-11)      & {\bf 5.7(-12)}  & 6.0(-12) \\
\hline
\label{tab:erfc}
\end{tabular}
\medskip{\protect\\
Notes.\protect\\
a(b) = a $\times$ 10$^b$. All the abundances are with respected to H nuclei.\\
Bold font indicates overestimation or underestimation by more than one order of magnitude compared with the observed values.\\
$^a$Observational data are taken from \citet{Vasyunin2013}.\\
$^b$Observational data are taken from \citet{Bacmann2016}.\\
$^c$Observational data are taken from \citet{Chacon-Tanarro2019},
assuming N(H$_2$) = 1.5 $\times$ 10$^{22}$ cm$^{-2}$ toward the ``methanol peak'' \citep{Jimenez-Serra2016}.\\
$^d$Observational data are taken from \citet{Jimenez-Serra2016}.\\
$^e$Observational data are taken from \citet{Vasyunin2017}.\\
$^f$Observational data are taken from \citet{Agundez2013}.\\
$^g$Observational data are taken from \citet{Cernicharo2020} assuming N(H$_2$) = 1.0 $\times$ 10$^{22}$ cm$^{-2}$.\\
$^h$Observational data are taken from \citet{Agundez2021}.\\
$^i$Observational data are taken from \citet{Tafalla2006}.\\
$^j$Observational data are taken from \citet{Jimenez-Serra2021} toward the ``methanol peak''.\\
}
\end{table*}

\section{Discussions and Conclusions}\label{sec:summary}
The MRN grain-size distribution \citep{MRN} is representative of the size of dust grains in diffuse clouds. 
As diffuse clouds condense to form dense clouds, grains may coagulate \citep{Chokshi1993}.
Consequently, the size-distribution of dust grains should vary with time. 
In this paper, following our previous work \citep{Chen2018} utilizing the MRN grain-size distribution,
we also adopt a grain-size distribution that considers grain coagulation and shattering \citep{HY09} to investigate 
their effect on the chemical evolution of dense clouds. 
Moreover, the radii of the smallest grains in this work are extended to around 1.5 $\times$ 10$^{-7}$ cm, 
thus, their temperature fluctuations due to the stochastic heating
are much larger than these in the previous work \citep{Chen2018}.
The stochastic heating of the grain is caused by the absorption of low energy (infrared wavelength) 
and high energy (cosmic-ray induced secondary FUV) photons.
Therefore, we also focus on studying the COMs formation in this work. 
The reactive desorption mechanism is included in models to elucidate 
how stochastic grain heating can help to explain the observed COMs in cold cores.

Speaking for the stochastic grain heating,
the contribution of the temperature fluctuations of the grains to the evolution of the chemistry should be significant.
As an example for model M2, the highest temperature spike that could be reached is 15, 30, and 49 K for
grains in bin 1, 2, and 3 (see Tab.~\ref{tab:sizeDistri} for the grain size) for the low energy photons heating, respectively.
While for the high energy photons heating, 
the highest temperature spike that could be reached is 20, 48, and 90 K for grains in bin 1, 2, and 3, respectively. 
The heating event for the low energy photons is approximately a few per 10$^{5}$ seconds for grains in bin 3,
while the heating event for the high energy photons is much rare, 
i.e., once per 10$^{9}$ seconds for a standard interstellar radiation field.
The time for the cooling of the grain down to 3 K is about 70000 seconds.
For larger grains, the heating events will be more frequent, but the fluctuations of the temperature will become smaller.
Therefore, compared with the standard grains used in the reference model, 
the moderately sized grains, i.e., grains in bin 1 and 2 in Tab.~\ref{tab:sizeDistri},
could stay in a higher temperature range for some durations, which will eventually affect the chemistry in the cold cores.

Our model results suggest that the total fractional abundances of granular species are not much affected by the grain coagulation.
Gas-phase CO$_2$ is one of the species whose abundances change significantly by grain coagulation.
Granular CO$_2$ are more likely to be formed on smaller grains 
with larger temperature fluctuations and then sublime quickly, so the abundances of gaseous CO$_2$ in models that 
consider grain coagulation are a factor of ten lower than that in models that adopt the MRN distributions. 
The effect of grain coagulation can affect some gas-phase species
by the number of small grains and their temperature fluctuations.
The most affected gas-phase species have the following characteristics: 
1) the contribution of their gas-phase production in the cells of large grains is not significant,
i.e~their abundances are relatively low at low gas-phase temperatures;
2) they are derived by the gas-phase chemistry in the cells of small grains, while the precursors are volatile 
and originate from the surface of these small grains.

It turns out that fewer surface species can be synthesized on the smallest grains whose radii are less than 2 $\times$ 10$^{-7}$ cm.  
There is less than one monolayer on these grain surfaces because many surface species can efficiently sublimate on these grains. 
The major granular species on the smallest grains are water and ammonia ice.
On the other hand, for grains larger than 2 $\times$ 10$^{-7}$ cm, 
the species accumulate on the grain surface, and ice mantles are developed,
with gH$_2$O, gCO, gCO$_2$, gH$_2$CO, gCH$_4$, gNH$_3$, and gCH$_3$OH as the major and most important constituents.
The ice compositions are enhanced for grains with the size of 1 $\times$ 10$^{-6}$ cm due to the temperature fluctuations.
When the size of the grain becomes smaller (~5 $\times$ 10$^{-7}$ cm), COMs begin to be efficiently produced.

The recombination of two radicals to form COM can only occur on grains whose radii are around 4.6 $\times$ 10$^{-7}$ cm 
so that their temperature fluctuations are moderate.
Moreover, methanol can also be efficiently formed on these grains. 
Photolysis of methanol, as well as hydrogenation reactions, can generate enough radicals. 
These radicals cannot sublimate but are able to diffuse in the grain mantles 
to recombine with each other to form COMs due to the moderate temperature fluctuations.
Therefore, models that include the stochastically heated grains could help to explain the observed COMs toward cold cores.

\section*{Acknowledgments}
We thank the reviewer for the helpful comments on improving the manuscript.
This research was supported by the National Natural Science Foundation of China (NSFC) grant No.~11988101 and No.~11725313.
Y.~W.~acknowledges the support by the NSFC grant Nos.~11973090, 11873094, 12041305, and the Natural Science Foundation of Jiangsu Province (Grants No.~BK20221163).
This research was carried out in part at the Xinjiang Astronomical Observatory.
The Taurus High Performance Computing system of Xinjiang Astronomical Observatory was used for the simulations.
We thank Hiroyuki Hirashita and Ji Xing Ge for sharing the HY09 data.

\section*{Data Availability}
The data underlying this article will be shared on reasonable request to the corresponding author.

%\appendix

\bsp

\label{lastpage}

\end{document}